\documentstyle[aaspp4,11pt,flushrt,twoside,times]{article}
\def\itm#1\par{\item#1\par}
\def\alwaysmath#1{\ifmmode{#1}\else{$#1$}\fi}
\def\ts{\thinspace}
\def\ngc#1{NGC\thinspace #1}
\def\etal{{\rm et al.}}
\def\msun{\alwaysmath{{M}_{\odot}}}

\def\teff{\alwaysmath{{T}_{\rm eff}}}
\def\logteff{\alwaysmath{\log{T}_{\rm eff}}}
\def\etar{\alwaysmath{\eta_{{}_{\scriptscriptstyle R}}}}
\def\feh{[{\rm Fe/H}]}
\def\ofe{[{\rm O/Fe}]}
\def\menv{\alwaysmath{M_{\rm env}}}
\def\Agbman{AGB-manqu\'{e}}\def\agbman{\Agbman}
\def\flashman{flash-manqu\'{e}}
\def\ocen{$\omega\,$Cen}
\def\etarhb{\alwaysmath{\etar(\rm RHB)}}
\def\etarr{\alwaysmath{\etar(\rm RR)}}
\def\etabhb{\alwaysmath{\etar(\rm 10000\,K)}}
\def\etafl{\alwaysmath{\etar(\rm flash)}}
\def\etapeel{\alwaysmath{\etar({\textstyle \rm peel-off})}}
\def\etaehb{\alwaysmath{\etar(\rm EHB)}}
\def\mcore{\alwaysmath{M_{{\rm c}}}}
\def\mflash{\alwaysmath{M_{{\rm fl}}}}
\begin{document}
\title{THE ORIGIN OF EXTREME HORIZONTAL \\ BRANCH STARS}
\author{Noella L. D'Cruz\altaffilmark{1}}
\authoremail{nld2n@virginia.edu}
\author{Ben Dorman\altaffilmark{1}$^,$\altaffilmark{2}}
\authoremail{dorman@shemesh.gsfc.nasa.gov}
\author{ Robert T. Rood\altaffilmark{1}}
\authoremail{rtr@virginia.edu}
\author{Robert W. O'Connell\altaffilmark{1}}
\affil{University of Virginia, P.O.~Box 3818, Charlottesville, VA 22903}
\authoremail{rwo@virginia.edu}
\altaffiltext{1}{nld2n@virginia.edu;~dorman@shemesh.gsfc.nasa.gov;~rtr@virginia.edu;~rwo@virginia.edu}
\altaffiltext{2}{Present Address:~Laboratory for Astronomy and Solar
Physics, Code 681, NASA Goddard Space Flight Center, Greenbelt, MA 20771}
\centerline{\it Submitted to the Astrophysical Journal}

\begin{abstract}
Strong mass loss on the red giant branch (RGB) can result in the
formation of extreme horizontal branch (EHB) stars. The EHB stars
spend most of their He core \& shell burning phase at high
temperatures and produce copious ultraviolet flux. They have very
small hydrogen envelopes and occupy a small range in mass. It has been
suggested that the ultraviolet excess phenomenon, seen in elliptical
galaxies and spiral bulges, is largely due to the ultraviolet flux
from EHB stars and their progeny.  The main motivation behind this
work is to understand how EHB stars can be produced over a wide range
of metallicities without fine tuning the mass loss process.

We have computed evolutionary RGB models with mass loss for stars with
main sequence ages between 12.5 and 14.5 Gyr, initial masses $< 1.1
M_\odot$ and metallicities of $\feh = -2.26,$ $-1.48, -0.78,$ 0.0, and
0.37. We used the Reimers formula to characterize mass loss, but
investigated a larger range of the mass loss efficiency parameter,
$\etar$, than is common. We adopted values of $\etar$ from 0 to 1.2 for
the compositions considered. Sufficiently rapid RGB mass loss causes
stars to ``peel-off'' the giant branch, evolve to high temperatures
and settle on the white dwarf cooling curve. Some such stars have
adequate helium core mass to undergo the helium flash, but only
later at high
temperatures: we call these ``hot He-flashers''.  After helium
ignition, the hot flashers lie at the very blue end of the horizontal
branch, forming a ``blue hook.''  The blue hook stars are a part of the EHB
population and  evolve into \agbman\ stars. The rest
of the peel-off stars form helium white dwarfs which we call
``\flashman'' stars. To understand how the number of EHB stars varies with
metallicity in a stellar population we considered how the zero-age
horizontal branch (ZAHB) is populated.  We assumed the distribution of
ZAHB stars to be driven by a distribution in the mass loss efficiency
(parametrized by $\etar$) rather than by a distribution in mass.  Our
results show that at low metallicity, EHB stars are produced if \etar\
is 2--3 times larger than the value producing normal ``mid-HB stars.''
However the range in \etar\ producing EHB stars is comparable to that
producing mid-HB stars.  Somewhat surprisingly, neither the range nor
magnitude of \etar\ producing EHB stars varies much metallicity.  In
contrast, the range of \etar\ producing mid-HB stars decreases with
increasing metallicity.  Hence the HB of populations with solar
metallicity and higher, such as expected in elliptical galaxies and
spiral bulges, will be bimodal if the distribution covers a
sufficiently large range in \etar.

\end{abstract}


\newpage
%----------------------INTRODUCTION------------------------------

\section{INTRODUCTION} \label{sec:intro}

Extreme horizontal branch (EHB) are stars which fail to reach the
thermally pulsing stage on the asymptotic giant branch (AGB) after
evolving off the zero-age horizontal branch (ZAHB). After core helium
exhaustion, EHB stars evolve into either Post-Early AGB (P-EAGB) stars
which leave the AGB before thermal pulsing or \agbman\ stars, which
never reach the AGB (Brocato, \etal\ 1990; Greggio \& Renzini 1990
[GR90]; Dorman, Rood, \& O'Connell 1993 [DRO93]). These stars have
undergone such extreme mass loss during their first ascent up the
giant branch that only a very thin hydrogen envelope survives. Stars
identified as EHB stars are found in low metallicity globular clusters
as an extension of the normal HB---perhaps most prominently in \ocen\
(Whitney \etal\ 1994) and \ngc{6752} (Buonanno \etal\ 1986;
Castellani, Degli'Innocenti, \& Pulone 1995). In higher metallicity
populations, EHB stars are found in the the metal rich cluster
\ngc{6791} (Liebert, Saffer, \& Green 1994) and as field sdB stars
(Saffer \& Liebert 1995; Dorman, O'Connell, \& Rood 1995 [DOR95]).
EHB stars and their \agbman\ and P-EAGB progeny are perhaps the prime
candidate objects for the far-ultraviolet upturn (UVX) observed in
elliptical galaxies and spiral galaxy bulges (Burstein \etal\ 1988;
GR90; Ferguson \& Davidsen 1993; DOR95).

One difficulty in understanding the EHB stars is that at a given
composition the range of total stellar mass which produces EHB
behavior is very narrow. Superficially, it appears that mass loss must
be ``fine tuned'' such that the red giants lose a large amount of
mass, but stop just short of losing the entire envelope. For globular
clusters where the blue HB is populated (the intermediate blue HB
(IBHB) in the nomenclature of DOR95) it is easy to imagine that some
EHB stars result from the high mass loss tail of the mass loss
distribution. At high metallicity as in \ngc{6791}\ ([Fe/H] $\sim
+0.2$) there appear to be EHB stars even though the IBHB is not
populated---Liebert \etal\ (1994) were unable to explain \ngc{6791}\
with a ``normal'' HB mass distribution.  Also, the sdB stars of the
Galactic field form a separate evolutionary group from the field blue
HB stars (Heber 1986) and make up a considerable proportion of all
UV-excess objects (Green, Schmidt, \& Liebert 1986).  Since the EHB
appears to occur over a wide range of metallicity, not only might the
mass loss require ``tuning,'' but that tuning might require
substantial adjustment with metallicity. Understanding how EHB stars
could be produced for a wide range of compositions, without fine
tuning, was the initial motivation for this investigation.

The effects of mass loss have been studied extensively in the
different phases of stellar evolution (see Dupree 1986 for a review).
If the sun is to be taken as an example,
low mass main sequence (MS) stars experience negligible mass loss,
$\sim 10^{-14} M_\odot\, {\rm yr}^{-1}$. If the rate of mass loss is
constant over the main sequence lifetime of a star then it loses on
the order of $10^{-4}\,M_\odot$ in about 10\ts Gyr. However, the mass
loss rate is expected to increase as stellar
luminosity increases and surface gravity decreases along the giant
branch. Both observations and empirical scaling laws such as the
Reimers' relation (Reimers 1975, 1977) suggest that a substantial
fraction of a star's mass could be lost on the giant branch (Dupree
1986). Since mass loss is a surface phenomenon, it only affects the
convective envelope of the RGB stars, and except in extreme cases,
evolution of the interior of the star continues without any
knowledge of the loss of the mass from the surface (see also
\cite{Jimenez95} and \cite{CC93} [CC93]). Because of this, mass loss
has ordinarily been neglected during red giant evolution, with the
total mass lost during this phase being removed at the helium flash for the
purposes of modeling later evolution.

However, with extreme, but plausible mass loss rates, a star can lose
so much mass that its total mass becomes comparable to its core mass.
The basic behavior of such stars is obvious (Renzini 1981; Caputo
\etal\ 1981).  At some point the envelope mass becomes so small that a
large convective envelope can no longer exist.  Then, instead of
continuing to the tip of the giant branch, the star ``peels-off'' the giant
branch and evolves towards higher temperatures at almost constant
luminosity.  We call this an RGB peel-off star.

There are two possible routes of evolution for peel-off stars. The
more obvious resembles the evolution of Post-AGB models (Sch\"{o}nberner
1979, 1983) but with the peel-off stars dying quietly as He white
dwarfs (He WDs). We refer to these as ``\flashman '' stars
(RGB-stragglers in the nomenclature of
\cite{Lur94} and Castellani, Luridiana, \& Romaniello 1994).
Initially, it was assumed that this was the only route followed by
peel-off stars. Hence, it was also assumed that the smallest envelope
mass that actually reaches the He-flash on the RGB, i.e.~the envelope
mass at peel-off, set the minimum envelope mass --- and thus the
maximum effective temperature --- for HB stars at the onset of He
burning.  Eventually CC93 found another
evolutionary path for peel-off stars. Despite the fact that the stars
were cooling along the He-white-dwarf cooling curve, for some range of
parameters a He core flash occurred. Even though the H-burning shell
was extinguished, the outer parts of He core were compressible enough
that contraction continued, leading eventually to helium ignition.  We
call these stars ``hot He-flashers'' to distinguish them from the
classic RGB tip flashers. Hot-flashers spend their entire He burning
life at high
\teff, first as extreme HB stars, and then as \agbman\ stars (CC93;
Castellani \etal\ 1995).

CC93 discovered hot-flashers in a study which incorporated mass loss
along the giant branch via the Reimers formula (Reimers 1977). From
CC93 as well as \cite{CLR94} and \cite{CDP95}, it was not clear
whether hot-flash behavior could be achieved without fine tuning the
Reimers mass loss efficiency \etar.  Our initial goal was to determine
the range of \etar\ leading to the high-temperature flashes as a
function of metallicity.  Anticipating that fine tuning would be
required, we would then explore other mass loss formalisms which would
more easily lead to the production of EHB stars. As it turned out,
fine tuning of \etar\ was not necessary. We expect that the hot-flash
phenomenon should occur for a fairly broad class of mass loss laws for
almost any composition.

We have computed RGB models with mass loss over a wide range of
metallicities. The calculations of our models, including our treatment
of mass loss are described in \S~\ref{sec:cal}. We present the results
of our RGB models and their corresponding HB models in
\S~\ref{subsec:RGBevol} and \S~\ref{subsec:HB} respectively. The
astrophysical significance of our results and the occurrence of EHB
stars in stellar populations with different metallicities is discussed
in
\S~\ref{sec:dis}. The conclusions are summarized in \S~\ref{sec:summary}.

%----------------------------CALCULATIONS-----------------------

\section {CALCULATIONS}	\label{sec:cal}

\subsection{Procedures} \label{subsec:procs}
We have computed evolutionary tracks from the zero-age main sequence
(ZAMS), using a modified version of the stellar evolution code STEV
(\cite{Dor92}). The details of the code can be found in Dorman (1992a)
and VandenBerg (1983, 1992).  Mass loss has been incorporated into the
code for stars with convective envelopes---the mass of the convective
envelope is reduced by an amount given by a specified mass-loss-law at
each timestep. A new internal mass mesh grid is created for each model
because of the reduced surface mass. Upon convergence of the stellar
structure equations, a valid stellar model of reduced mass is
obtained.

We neglect mass loss along the main sequence. For our evolutionary
tracks we assume that mass loss takes place only during the red giant
branch phase. The mass loss prescription that we use here is the
widely used Reimers Formula (Reimers 1975, 1977):

$$ \dot{M} = -4 \times 10^{-13}\: \etar\:\frac {L}{g R} \:\: M_\odot\,
{\rm yr}^{-1}$$

\noindent where $\etar$ is a mass loss efficiency parameter, $L$ is the
luminosity, $g$ is the surface gravity,
and $R$ is the radius, with $L,~g,~ {\rm and}~R$ in solar units.
To produce a ``normal''
globular cluster horizontal branch requires $\etar$ $\sim 0.25 - 0.5$
(\cite{Renzini81}).

If the mass loss rate is large enough, the star leaves the RGB before
reaching the tip. It evolves toward higher temperatures at constant
luminosity after RGB peel-off. This results in a decrease in radius
and an increase in surface gravity, which should cause the mass loss
rate to decrease.  For hot stars, different mass loss mechanisms may
be important (Chiosi \& Maeder 1986), but the mass loss rates are
expected to be significantly smaller in the tightly bound remnants of
low mass stars than in young massive stars in which high temperature
winds have been observed and modeled.  In anticipation of this
physically expected decrease in mass loss rate, we turn off the mass
loss after the convective envelope contains less than $10^{-3}
M_\odot$.

The code STEV calculates stellar models using the Eggleton
non-Lagrangian scheme (\cite{Eggleton71}) for the conventional red
giant branch evolution (\cite{VdB92}), and a standard Lagrangian grid
for all other evolutionary phases.  The non-Lagrangian scheme was
designed primarily for the evolution of stars with a thin burning
shell, as an alternative to the shell-shifting technique
(see Sweigart \& Gross 1978).
It provides a faster means of computing the evolution when
there is a rapidly advancing shell, since it does not require that the
timestep be small enough to ensure that only a small mass fraction of
the burning shell is consumed between models.
Typical RGB
calculations require about 600 models, a third of which are used to
compute small timesteps close to the He-flash.
However, since our implementation of the Eggleton scheme provides only
limited capability for following a detailed nuclear reaction network
(but see Han, Podsiadlowski, \& Eggleton 1995), the calculations are
halted after the star leaves the RGB, and continued with the
Lagrangian code, which follows 7 elements in a fully implicit scheme.
The non-Lagrangian scheme computes the H, He and $^{14}$N profiles,
but is able to follow the evolution of the other elements in a
separate calculation, for use at the transition.  The conversion from
one evolutionary scheme into another does not affect the qualitative
behavior of the models.

The surface boundary condition is computed differently for the three
sub-solar metallicities and the other two metallicities. The code has
the option of using surface pressures derived from computed model
atmospheres (\cite{VB85}) or assuming a T-$\tau$ (temperature
stratification) relation. Since the model atmospheres are available
only in a restricted range in temperature, evolutionary tracks with
wide variations in temperature and gravity cannot use the model grids
for the entire evolution.  We chose to use the model pressures for
$\feh < 0$ until the tracks peeled off the RGB, after which we used
the Lagrangian scheme and the Eddington approximation.  The result is
a gap in the evolutionary sequences in the more metal poor
calculations at the transition point.  For higher metallicity models
the same procedure cannot be used since the pressure grids, where
available, do not cover the range of surface values during the crucial
period of rapid mass loss.  In this case, the Eddington approximation
was used from the ZAMS. The stellar models were converted from
non-Lagrangian co-ordinates to Lagrangian co-ordinates at which point
$\menv \approx 10^{-5}\, M_\odot$ after peel-off.

For tip-flashers evolution was continued until $L_{\rm He} \geq 62.5
L_\odot$. In the case of the hot flashers and the \flashman s the code
ran until it was no longer able to continue the evolution. This
occurred at $L_{\rm He} \ga 30 L_{\rm H-shell}$ for the hot flashers.
Instead of pursuing the lengthy computations required to go through
the He flash, the stars were assumed to appear on a ZAHB with varying
core mass.  ZAHB models were created using the initialization
procedure outlined in Dorman, Lee, \& VandenBerg (1991), with the
hydrogen shell composition profile being extracted from the last
converged pre-flash model for each star. Since the stars which lose
more mass have different (thinner) hydrogen burning shells, the
relevant composition profile was found for each value of $\etar$ for
which the model underwent the helium flash.

\subsection{RGB Evolution} \label{subsec:RGBevol}

	We computed a number of RGB models with different values of
Reimers' $\etar$ for [Fe/H] = $-2.26, \:-1.48,$ $\: -0.78, \: 0.0,$
and 0.37. We chose the initial mass of the stars such that their MS ages
were between 12.5 and 14.5\ts Gyr. Thus, in our discussion below we
will be considering different stellar populations at roughly constant
age. At each metallicity, we calculated RGB models for $\etar = 0.0,
0.3, 0.6$. We increased $\etar$ in steps of 0.05 for $\etar > 0.6$
until there were at least two He-WD tracks for each metallicity. A
summary of the calculations for all the metallicities is given in
Table 1.
\begin{figure}[htbp]
\vbox to8.0in{\rule{0pt}{3in}}
\includegraphics{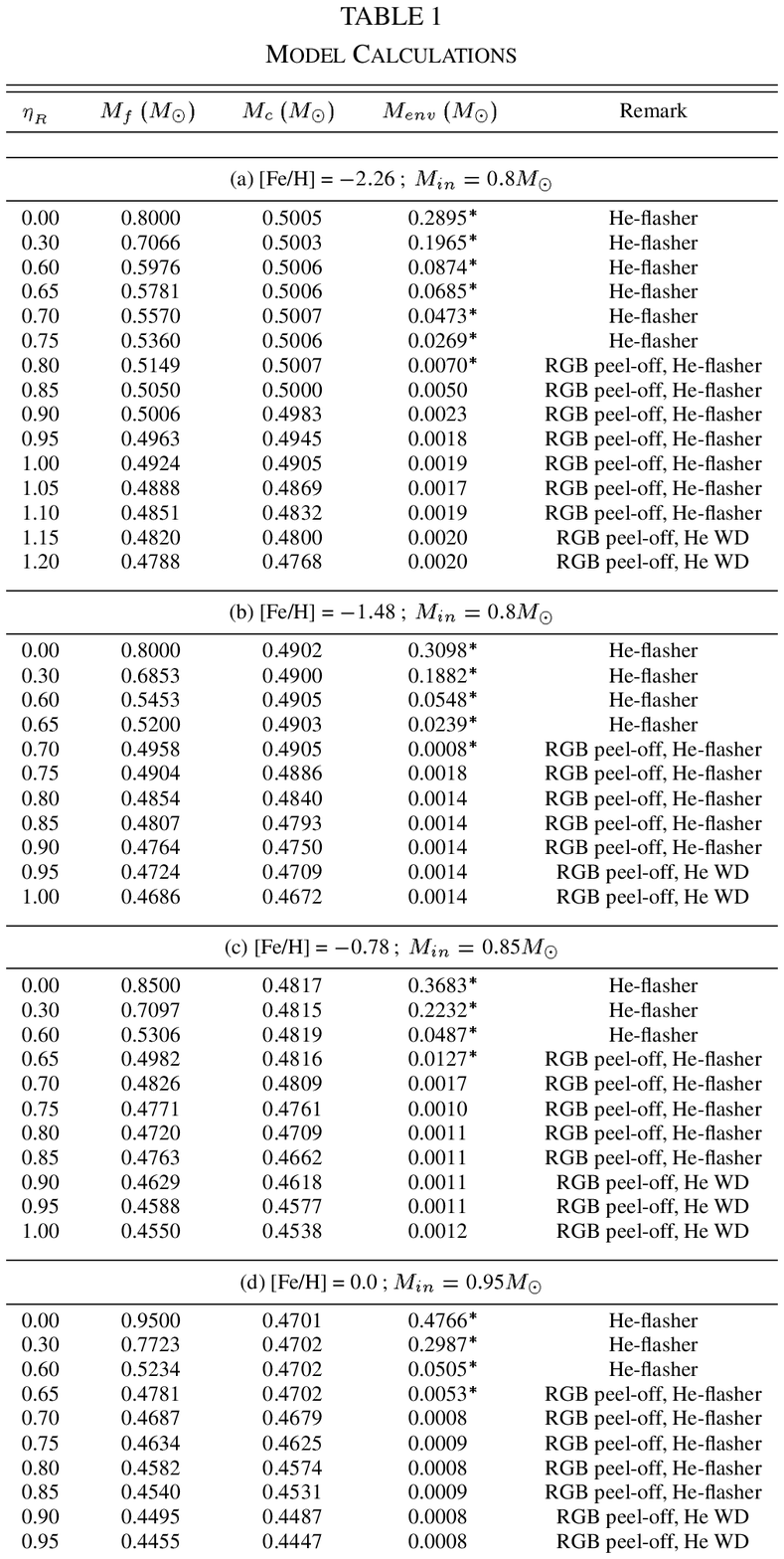}
\end{figure}
\begin{figure}[htbp]
\vbox to3.5in{\rule{0pt}{3in}}
\includegraphics{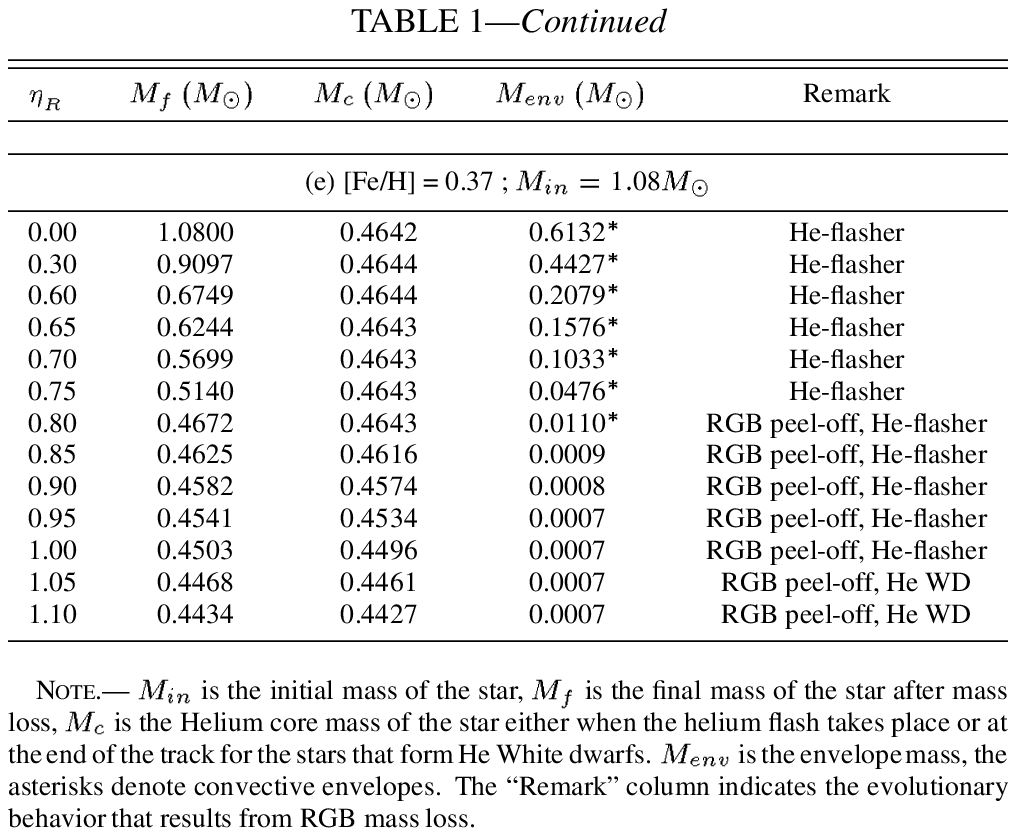}
\end{figure}
All models of the same metallicity are grouped together, with
the metallicity and initial stellar mass, $M_{in}$, indicated at the
beginning of each group. The first column lists the values of $\etar$
for which RGB models were created.  The second and third columns
contain the final mass, $M_f$, and the He-core mass, $M_c$, of the stars
respectively.  The final envelope masses, $\menv$, of the RGB models
are given in the fourth column. The asterisks denote convective
envelopes, the rest of the envelope masses being the difference
between the core mass and the total stellar mass.  The fifth column,
titled ``Remark'', indicates the evolutionary behavior of the models
i.e.~a star marked as ``He-flasher'' is a classic RGB tip flasher,
while a remark of ``RGB peel-off, He flasher'' indicates a hot
He-flasher, and ``RGB peel-off, He WD'' means that the star peeled-off
the RGB and failed to ignite helium.

	The HR diagram of the RGB tracks for the models listed in
Table 1 are shown in Figure~\ref{fig:RGB}.
\begin{figure}[htbp]
\vbox to8.0in{\rule{0pt}{3in}}
\includegraphics{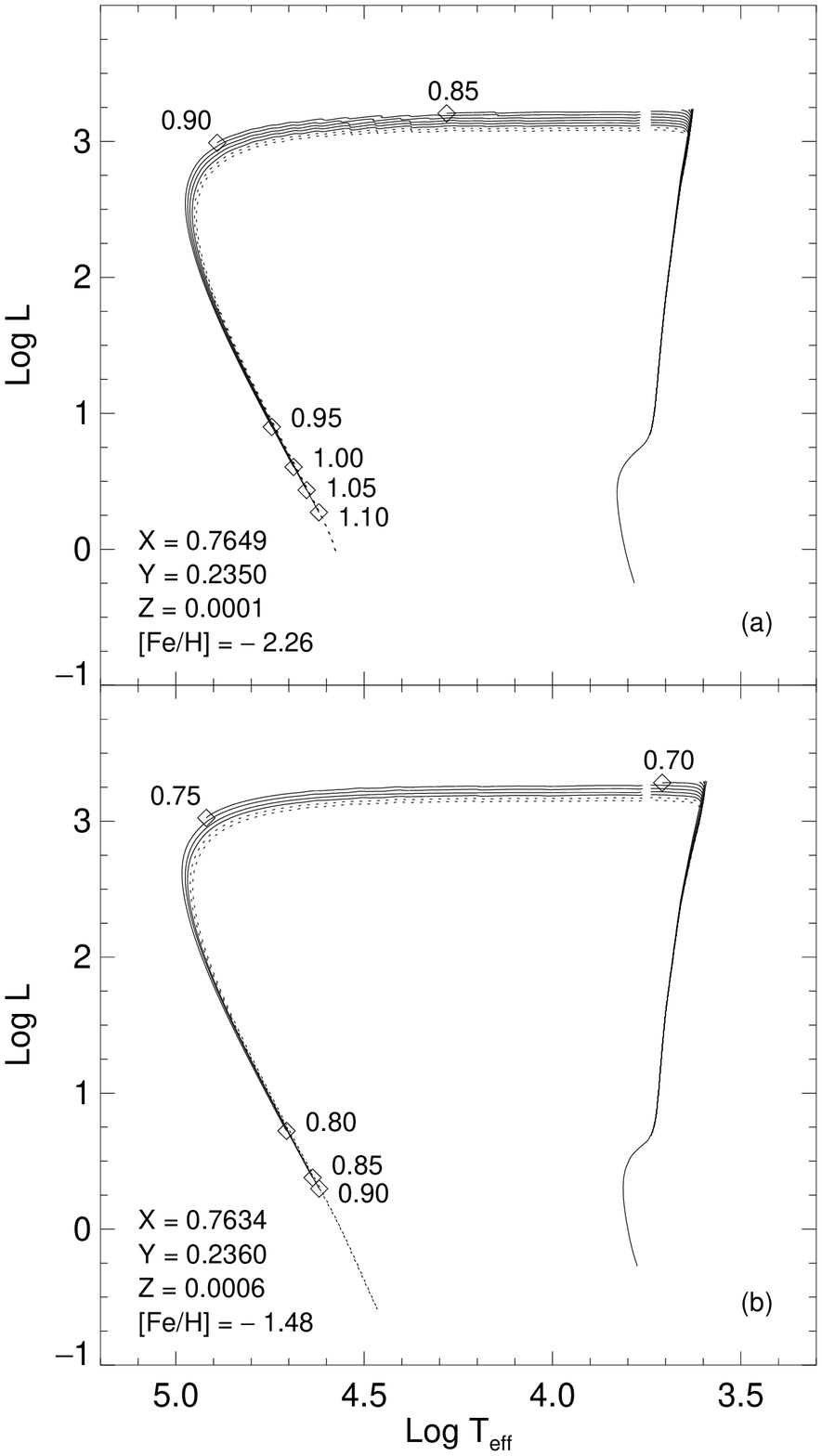}
\end{figure}
\begin{figure}[htbp]
\vbox to8.0in{\rule{0pt}{3in}}
\includegraphics{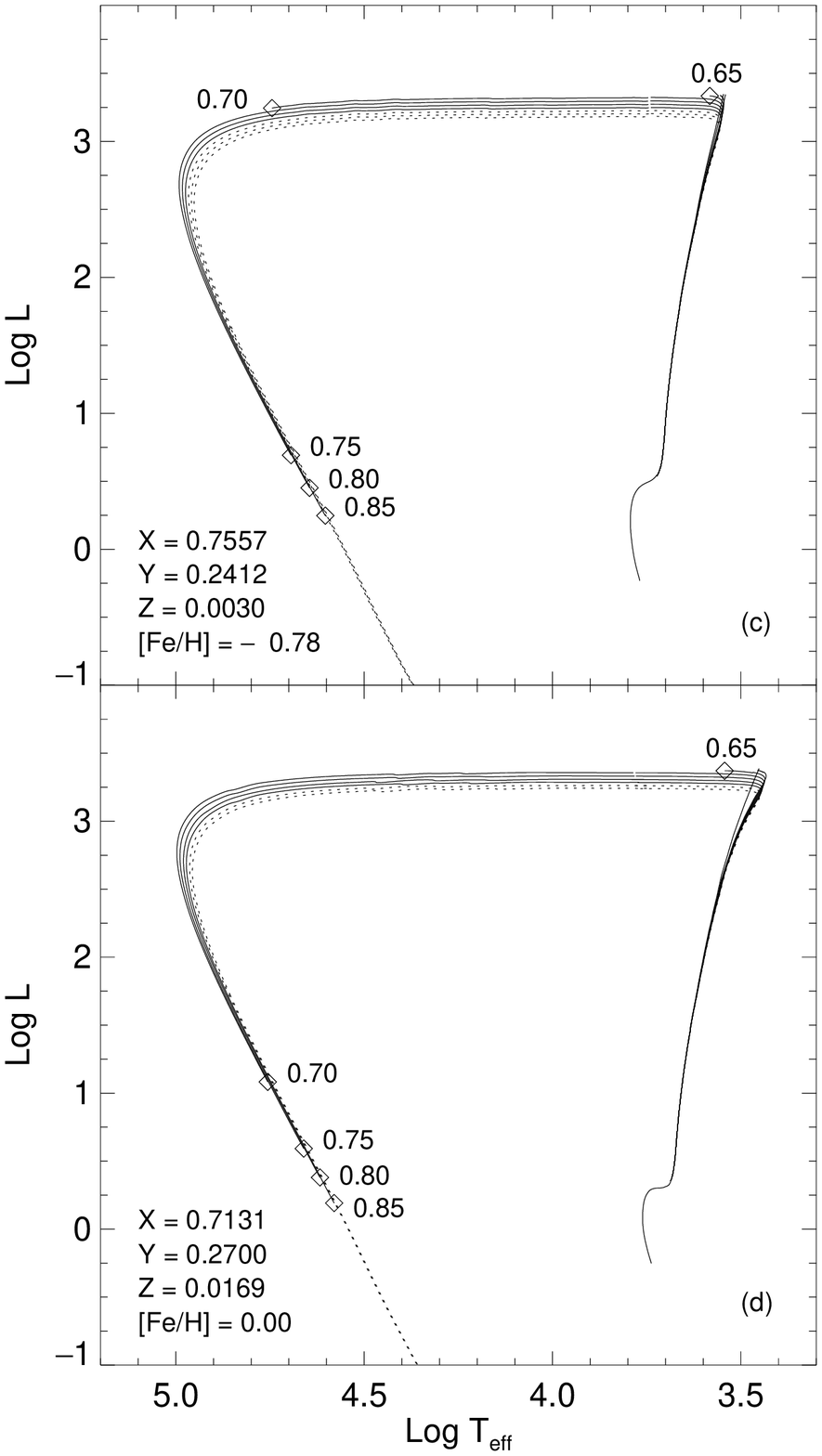}
\end{figure}
\begin{figure}[htbp]
\vbox to5.0in{\rule{0pt}{3in}}
\includegraphics{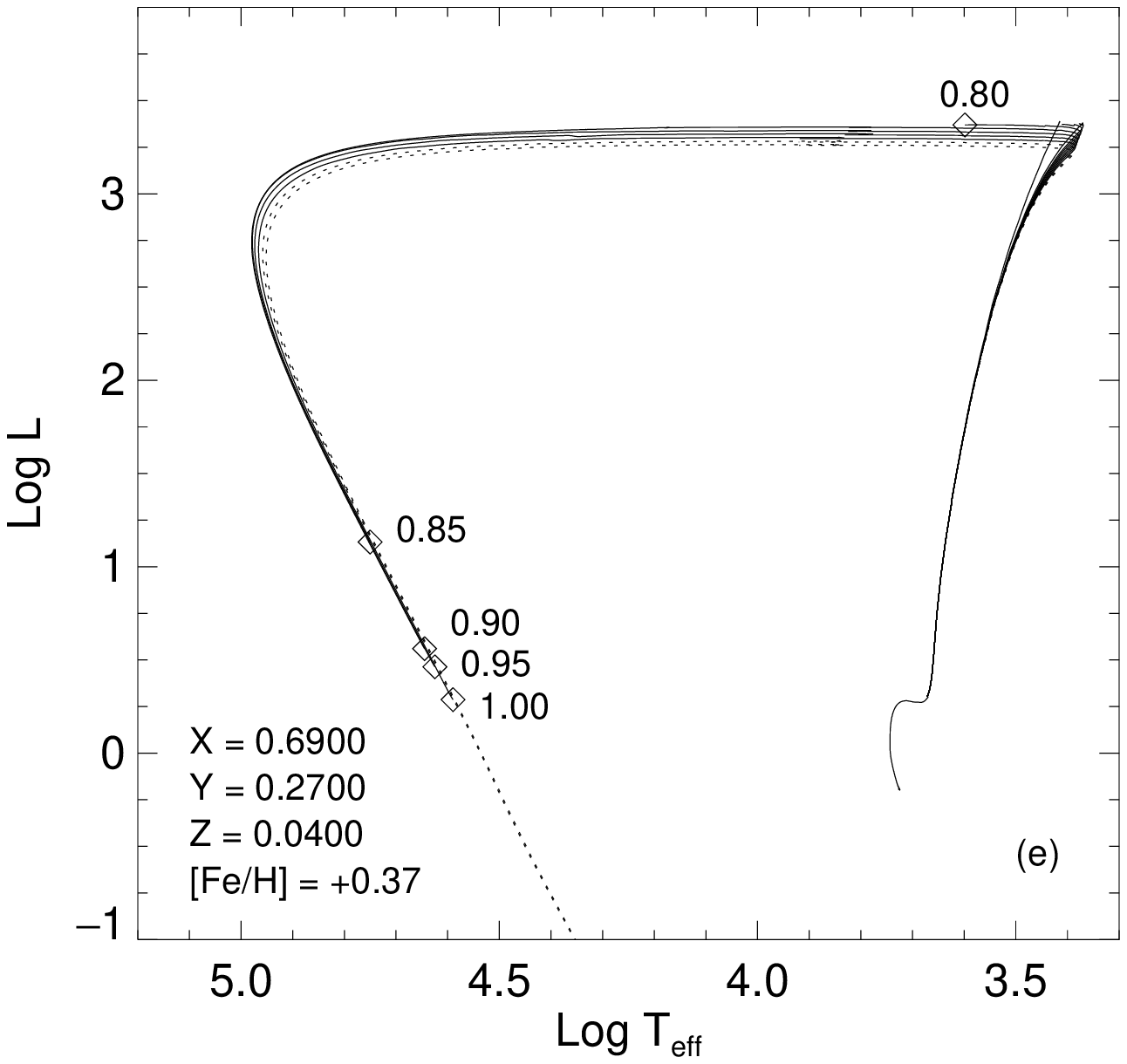}
\caption[RGB Tracks]{\small HR Diagram of the red giant branch evolutionary
tracks with mass loss for the metallicities considered--- $\feh =
-2.26, -1.48, -0.78, 0.0, 0.37$. The hydrogen abundance, $X$, the
helium abundance, $Y$, the heavy element abundance, $Z$, and the
metallicity, [Fe/H] are shown for each panel. The solid lines are
tracks which end with a helium flash. For the RGB peel-off stars the
location of the helium flash is indicated by an open diamond and the
corresponding $\etar$ value is listed next to the symbol.  These stars
are the ``Hot - He flashers.'' The dotted lines are tracks which form
He White Dwarfs. The break in the peel-off tracks is where the
conversion from non-Lagrangian to Lagrangian evolutionary schemes
takes place. Table 1 details the model parameters.\label{fig:RGB}}
\end{figure}
 The figure shows the
evolution of the models from the ZAMS until helium flash or the He WD
phase. All five metallicities are shown. The solid lines are the
tracks of flashing models, the dotted ones are \flashman\ models. The
location of the helium flash for stars that flash away from the RGB
are indicated by open diamonds and have the corresponding $\etar$
value listed next to them. The break in the peel-off tracks is where
the conversion from non-Lagrangian to Lagrangian co-ordinates takes
place. We see from Figure~\ref{fig:RGB} and Table 1 that there
is a range of masses for which models peel-off the RGB but still
undergo the helium flash for all metallicities. The hot-flash behavior
occurs for values of \etar\ 2--3 times larger than that required to
produce normal HB stars. Contrary to our initial expectations,
hot-flash behavior occurs for a fairly large range of \etar. Some of
the peel-off stars show weak hydrogen shell flashes while moving from
the RGB to the cooling curve.  This is most clearly seen in the $\feh
= -2.26$ tracks in Figure~\ref{fig:RGB}(a). Whether these flashes are
numerical or due to a real instability is not known. The models of
CC93 do not show similar flashes.

	None of our low metallicity models have enhanced oxygen.
Enhancement of oxygen and other $\alpha$-nuclei elements is often
found in metal poor stars (see \cite{WST89} for a review).  Increasing
oxygen affects RGB evolution in two ways: the rate of evolution is
sped up (at a given luminosity) and the flash occurs at higher
luminosity. The track in the HR diagram is essentially unshifted,
because the opacity at low $T$ is dominated by other elements (Renzini
1977).  Because of the faster evolution the core mass at flash is
smaller.  As far as mass loss goes there are compensating factors:
there is less time to lose mass as the evolution is faster, but the
evolution to higher $L$ increases mass loss. We performed one
comparison case with $\feh = -2.26$, $\ofe = 0.0, 0.75$ and $\etar =
0.0, 0.95$. As expected the oxygen enhanced models have smaller core
masses for both $\etar$ values and the flash occurs at higher $L$
for the $\etar = 0.0$ case. For $\etar = 0.95$, more mass was lost and
the star peels off at higher $L$ compared to the $\ofe = 0$ case.
Since our results suggest no change in qualitative behavior with total
metallicity we expect the effect of enhanced O to be small.

As noted earlier, the mass loss was terminated when the envelope mass
becomes less than $10^{-3} \,M_\odot$. To see whether this assumption
affects the hot-flash behavior, we also computed a track in which the
mass loss was allowed to continue until the envelope mass was so small
that it would be entirely removed within a timestep.  The core mass of
the star in this case changed by $10^{-3} M_\odot$, which is of the
order of the numerical uncertainty in determining the core mass at
the He-flash.
This again suggests that the arbitrariness of the point at which
mass loss was terminated does not change the
qualitative behavior of the models.

One particularly important result to realize here is that the \mcore\
for the hot-flashers is smaller than the (more or less) constant value
for the tip-flashers and that it decreases as \etar\ increases. It is
commonly believed that the rising temperatures in the He core are due
to the march of the hydrogen burning shell through the star.  Our
results show that the core temperatures rise because of the
contraction of the core, and that contraction continues whether or not
the mass of the core is increasing. We find that any star which has a
surviving H-envelope when its He core is within 0.015--0.018\msun\ of
\mflash, the classical core mass at flash, will undergo a He
flash. On the RGB near the flash $L \sim \mcore^5$, so any star which
gets to within $\sim 0.4\,$mag of the tip luminosity will flash either
at the tip or as a hot-flasher.  {\it Hence any mass loss law that
produces a substantial fraction of the total mass lost in the upper
$\sim 0.4\,$mag of the RGB has the potential to produce a substantial
population of hot-flashers.} This result, as well as that of
Castellani \etal\ (1995) implies that there is no minimum envelope
mass for the HB. This finding contradicts the assumption of
\cite{Yi95} that He burning cannot be initiated for $\menv \leq 0.02
\msun.$ \cite{Yi95} seem to have based their assumption on the work of
Sweigart, Mengel, \& Demarque (1974), who used a plausibility argument
to suggest that the probability of having HB stars with $\menv \leq
0.02 \msun$ is small.

The peel-off stars will be indistinguishable from P-EAGB stars as
they evolve to the cooling curve before the helium flash. Their
luminosity ($L \gtrsim 1000\,L_\odot$) and evolutionary timescale when
crossing the HR diagram (1--2 Myr) are similar.  Both types of objects
would be rare in populations as small as globular clusters because of
the brevity of this crossing phase.  In a cluster containing 10$^3$
RGB stars brighter than the HB and with lifetimes of $10^8\,$yr, one
would expect to find $\sim 10f_{\rm peel-off}$ peel-off stars, where
$f_{\rm peel-off}$ is the fraction of RGB stars that peel-off the RGB.
The value of $f_{\rm peel-off}$ must be less than the fraction
$N(EHB)/N(HB)$. The globular cluster $\omega$ Cen contains the largest
fraction of EHB stars at roughly 20\% [DOR95]. For this cluster one
would expect to have one or no RGB peel-off stars. This is consistent
with the number of stars predicted to exist at this luminosity. Since
these stars are so rare and are similar to P-EAGB stars, it would be
very difficult to verify their existence as such in the Galactic field
or in integrated light.

\subsection{Zero-Age Horizontal Branch Models and post-RGB evolution}
\label{subsec:HB}

ZAHB models for each composition are shown in Figure~\ref{fig:ZAHB}.
\begin{figure}[htbp]
\vbox to8.0in{\rule{0pt}{3in}}
\includegraphics{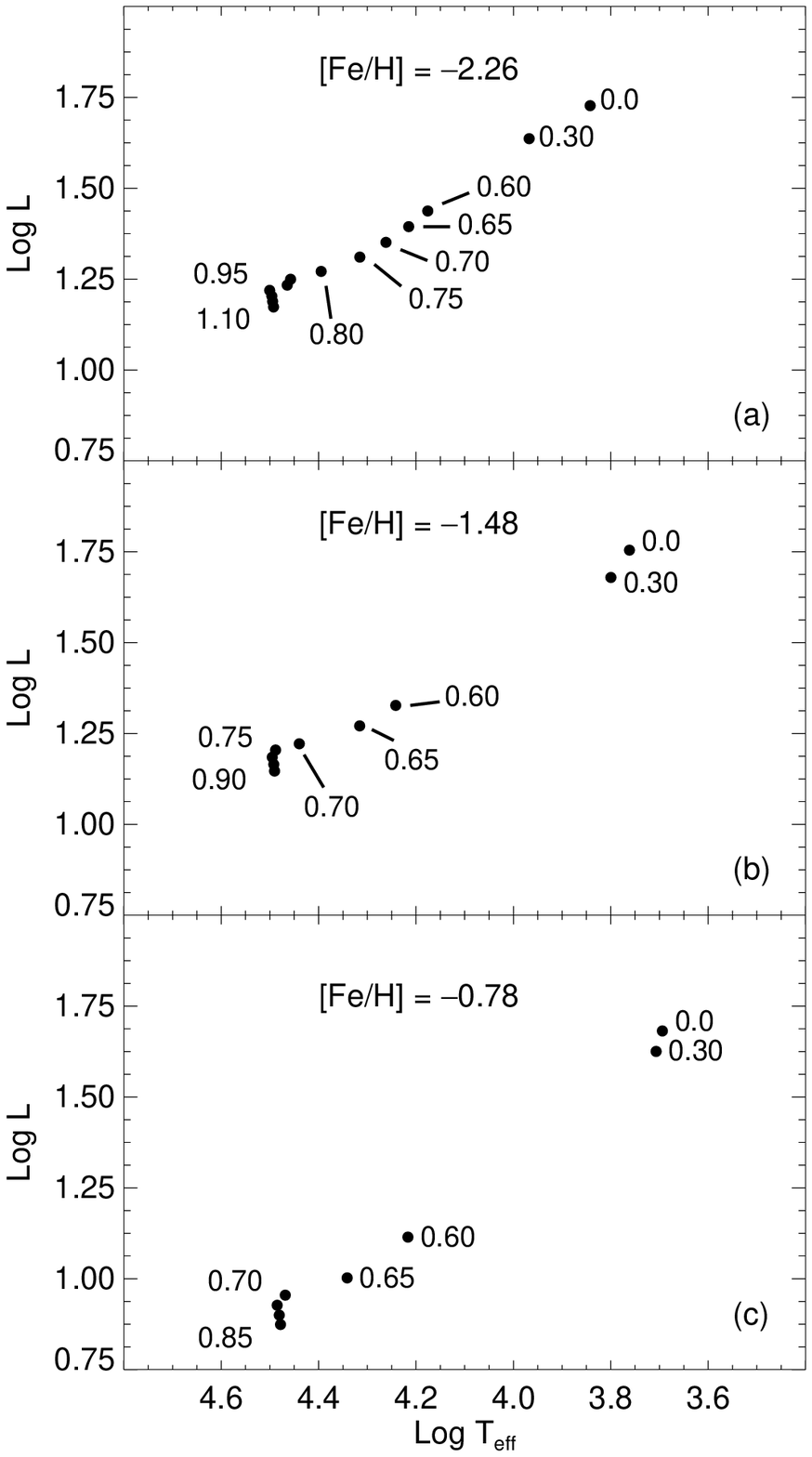}
\end{figure}
\begin{figure}[htbp]
\vbox to5.5in{\rule{0pt}{3in}}
\includegraphics{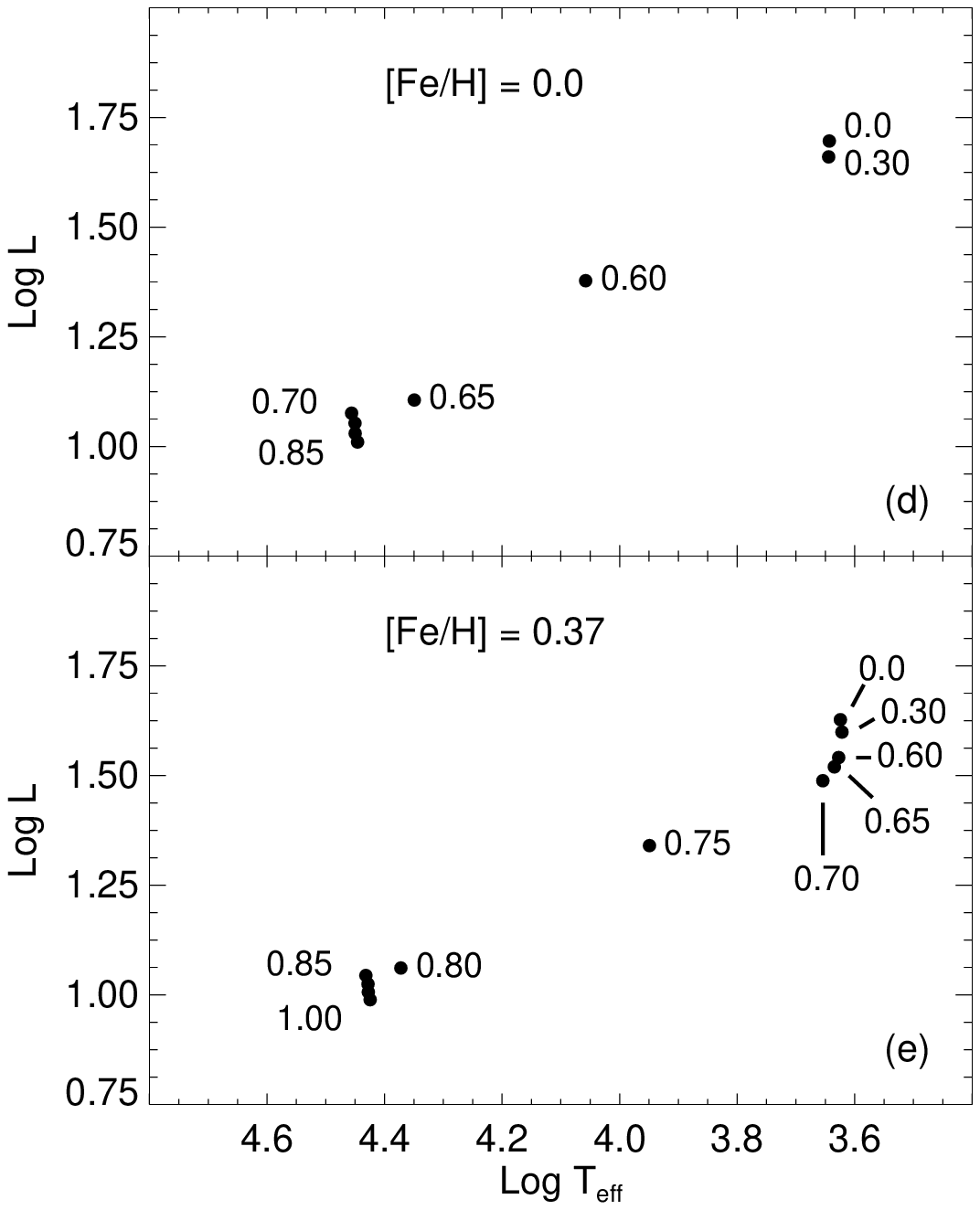}
\caption[ZAHB Models]{\small The HR diagram of ZAHB models corresponding
to the flashing RGB models from
Figure~\ref{fig:RGB} are shown for all
five metallicities. The corresponding $\etar$ values are marked.
The hot He-flashers form a blue hook at the blue end of the ZAHB at
all metallicities. \label{fig:ZAHB} }
\end{figure}
The hot-flashers do not follow the classical ZAHB as defined by Iben
\& Rood (1970) which is a constant core mass sequence. Their helium
core mass is less than that of the normal ZAHB models. These stars
have very small envelopes and hence will be very blue, and slightly
less luminous than the regular ZAHB.  They form a ``blue hook'' at the
blue end of the ZAHB as can be seen in Figure~\ref{fig:ZAHB}. The blue
hook stars are a part of the EHB population and have almost the same
envelope masses but different helium core masses.  Therefore they form
a sequence that is fairly constant in temperature but has a spread in
luminosity.  The temperature spread is $\leq 750 K$.  The length of
the hook can be estimated using $L \sim L_{\rm He} \sim M_{\rm
core}^3$ to be $\sim 0.1\,$mag. The blue hook sequence is cooler by
$\Delta \logteff \sim 0.1$ than the helium main sequence (Castellani
\etal\ 1995).

Blue-hook stars will evolve into \Agbman\ objects with similar
behavior to that of the
lowest mass ``traditional'' ZAHB models of DRO93. Both the blue-hook
EHB stars and their \agbman\ progeny will emit considerable UV
radiation. We computed the post-RGB evolution of two such models for
each metallicity. The two models that were followed had $\etar =
\etafl$ and $\etapeel +0.05$, where $\etafl$ is the highest value of
$\etar$ for which a helium flash occurs and $\etapeel$ corresponds to
the transition between classic red giant tip flashers and hot
He-flashers. The rest of the ZAHB stars are expected to evolve as
normal (see DRO93).  Figure \ref{fig:AGBMan} shows the post-RGB
evolutionary tracks of the two blue hook stars whose evolution we
computed.
\begin{figure}[htbp]
\vbox to7.0in{\rule{0pt}{3in}}
\includegraphics{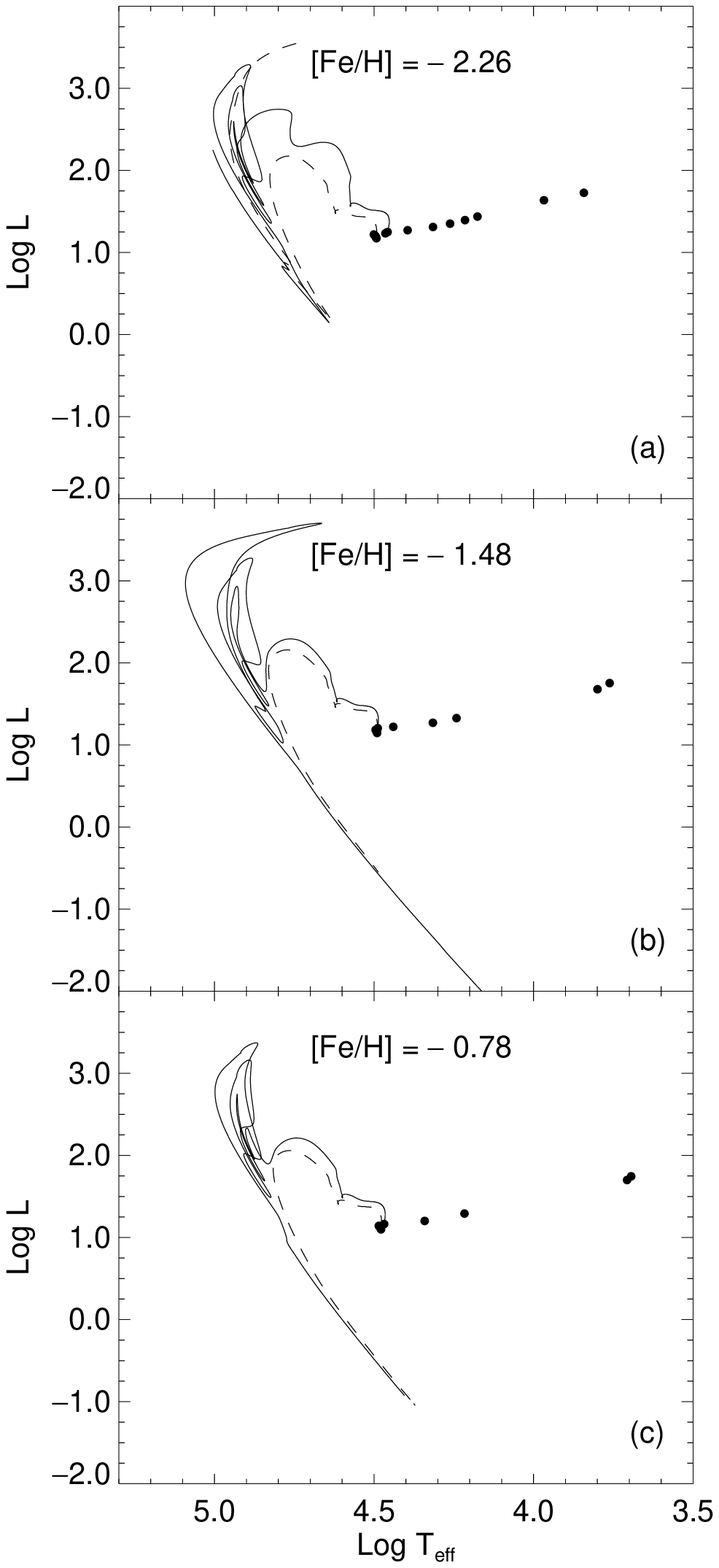}
\end{figure}
\begin{figure}[htbp]
\vbox to6.0in{\rule{0pt}{3in}}
\includegraphics{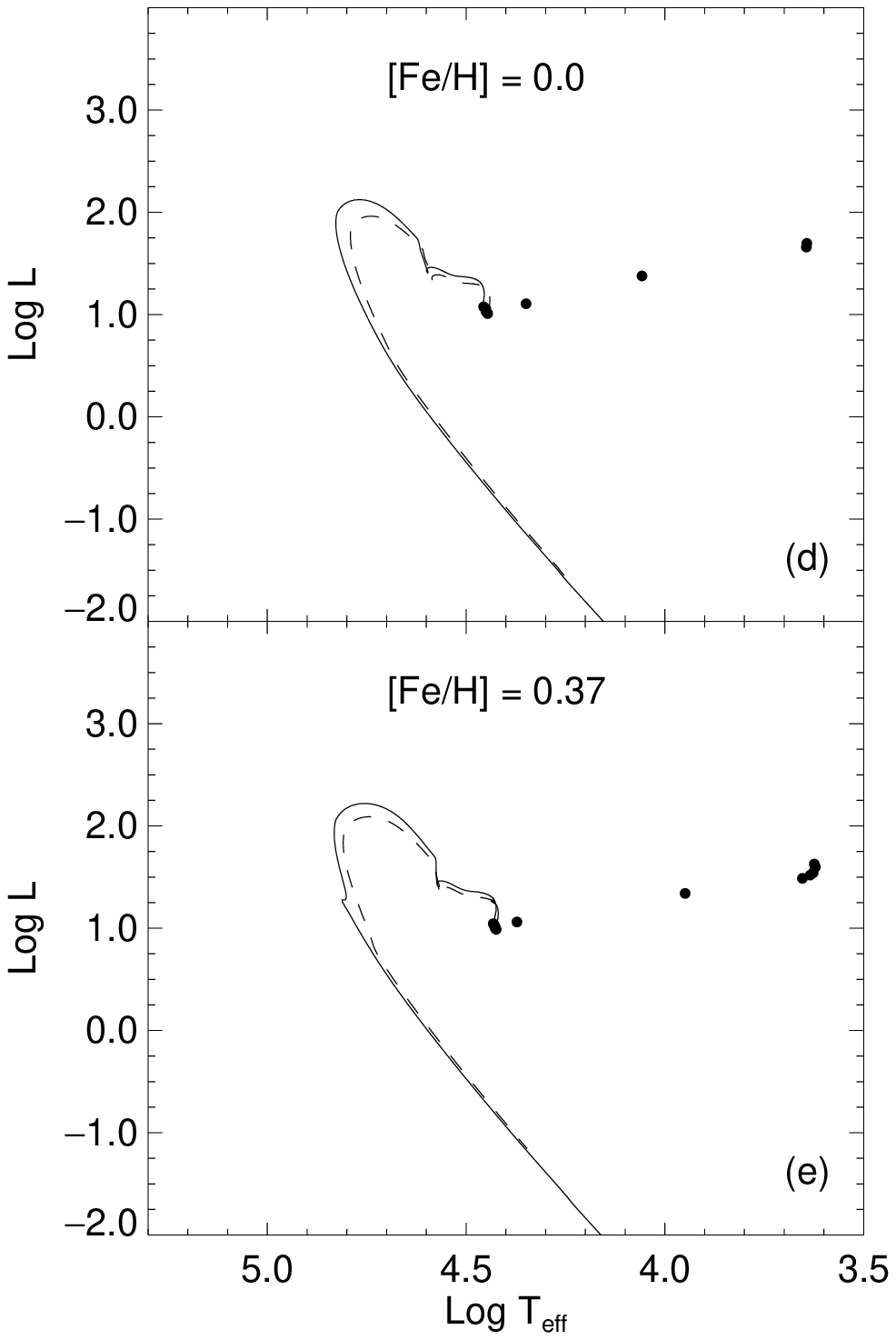}
\caption[AGB-Man]{\small The ZAHB Models are shown in this figure with
two \agbman\ tracks for all metallicities. The
\agbman\ tracks are for two stars in the blue hook as indicated
in the text. \label{fig:AGBMan}}
\end{figure}
For the initial vertical part of the tracks ($\sim 1\:
$mag), evolution is relatively slow and accounts for about 80\% of the
post-RGB life time.  An ensemble of blue-hook stars will form an
almost vertical band of hot HB stars which starts about 0.1 magnitudes
below the ZAHB with $\sim 80$\% of the stars lying within about one
magnitude above it.  With some imagination one can see such a sequence
in the hot HB stars observed in \ocen\ (Fig.~10 of \cite{Whit94}).

%-----------------------DISCUSSION--------------------

\section{DISCUSSION} \label{sec:dis}

Our use of Reimers' formula as a mass loss prescription could be
questioned. This formula has no physical basis and was constructed on
the basis of observation and dimensional analysis. There are alternate ways of
parametrizing mass loss, none of which have been found to be
exceptionally better or worse than Reimers' formula (see Dupree 1986;
Chiosi \& Maeder 1986). Ideally, one would like to use a mass loss
rate that is derived from a mass loss mechanism believed to operate in
cool giants. Unfortunately the exact mechanism that drives mass loss
is not known; the gross stellar parameters (temperature, luminosity,
gravity, metallicity) are obviously insufficient to determine the mass
loss uniquely.  Instead, the inferred spread in total mass lost may
somehow be tied to individual properties such as rotation rates (Fusi
Pecci \& Renzini 1978) and magnetic field strengths, and may be related to
the abundance variations detected in globular cluster stars (see Kraft
1994 for a review).

As for possible physical mechanisms for mass loss, thermally driven
winds, wave-driven winds, radiation pressure on molecules and/or dust
grains, and winds resulting from shock waves are those most widely
cited (see Holzer \& MacGregor 1985 for a review). It is possible
that more than one mechanism operates in a single star simultaneously.
For the Sun, it is now believed that the effects of rotation and
convection lead to the magneto-hydrodynamic heating of the solar
corona, which gives rise to the solar wind (Spruit, Nordlund, \& Title
1990).

Observations of cool, low-gravity stars show that the outflowing
material is moving with asymptotic velocities that are considerably
lower than the escape velocity of the star. Therefore, only a small
fraction of the energy that is added to the stellar atmosphere for the
purpose of fueling the wind is used to lift the wind out of the
gravitational field of the star. From the low asymptotic velocities
one can actually expect to learn very little about the mechanism
driving the mass loss since we observe only a small part of the
driving energy (Holzer \& MacGregor 1985). The electron temperatures
in cool supergiants are found to be around 10$^4\,{\rm K}$ as opposed
to 10$^6\,{\rm K}$ for the solar corona, which rules out thermally
driven winds (Dupree 1986). Large rates of mass loss also imply that
the mass loss mechanism is fairly efficient close to the stellar
surface where the densities are relatively high, which in turn places
a constraint on the temperature at which the mechanism operates.

Observations show that mass loss rates increase with decreasing
surface gravity and increasing stellar luminosity. The mass loss rates
themselves are uncertain in some cases to a factor of 5 or even 10.
Obtaining mass loss rates is a complex process and only a few stars
have well determined mass loss rates. The spectral lines (Ca H \&
K, H$\alpha$, etc) used to study mass loss originate at different
depths in the stellar atmosphere. Hence a better understanding of the
atmospheric structure and the origin of the lines is necessary. In the
face of our ignorance on this crucial matter, the Reimers formula
seems as good a parametrization as any, in that it obeys the observed
functional dependences and is dimensionally correct.

The conclusions we draw depend mainly on the escalation of the assumed
mass loss rate close to the RGB tip. Our generic results for core He
burning and beyond should still hold if mass loss rates increase close
to the RGB tip at least as sharply as in the Reimers formula.  Using
the Reimers formula or any other formulation which depends only on
surface parameters, mass loss continues until the star peels off the
RGB. These stars undergo He ignition so long as peel-off occurs
sufficiently near ($\sim0.4\,$mag) the RGB-tip. The helium cores are
slightly smaller than those produced in RGB tip-flashers, and the
ensuing evolution is the hot-flash/blue-hook behavior described here.
We anticipate that a physically motivated mass loss law will almost
certainly depend on the global properties of the convective zone. In
reality mass loss may stop as the convective envelope approaches some
minimum size $M_{\rm ce,{min}}$. After mass loss is terminated $M_{\rm
env}$ will continue to decrease as the H burning shell advances.
These stars will still ignite He if the mass loss is turned off within
$\sim 0.4\,$mag of the tip. Depending on the circumstances, He ignition
could be on the RGB or in a hotter phase. The core He burning stars
resulting from these stars would have envelopes ranging from our small
values up to $M_{\rm ce,{min}}$. The blue-hook will be smeared in
\logteff.  Indeed, the observed degree of smearing of EHB stars can
give a clue to the mass loss mechanism.

The ``standard model'' of the HB requires that individual RGB stars
arrive on the ZAHB with a distribution $P(M_{\rm ZAHB})$ of masses
(Rood 1973). This distribution together with the metallicity gives
rise to the different HB morphologies observed in globular clusters.
The distribution has typically been modeled by a gaussian truncated in
some way to account for physical limits to the mass (Rood
1973)\footnote{There are subtle differences in the truncation schemes
of the various workers which could affect extremes like EHB
populations.}.  {\it There was no physical basis for selecting the
form of $P(M_{\rm ZAHB})$ used}, other than that it seemed to work and
that random variables often tend toward being gaussian (i.e.~the
Central Limit Theorem). Indeed there are some counter examples where
this kind of formalism fails (e.g., Rood \etal\ 1993; Liebert \etal\
1994).  The accuracy of the assumption is hard to assess since
existing data do not allow us to distinguish easily between different
intrinsic mass distributions.  Below we consider the ZAHB population
to be driven by a function of $P(\etar)$ instead of $P(M_{\rm ZAHB}$).
As far as matching globular cluster HBs, the HB mass depends almost
linearly on $\etar$ over most of the HB for appropriate compositions.
Thus any distribution in $M_{\rm ZAHB}$ can be mapped into a
distribution in
\etar.  In this scheme, the star-to-star variation in \etar\ is a way
of modeling the effects of individual differences among stars that
cause the mass dispersion. One might think of the parameter $\etar =$
\hbox{${\etar}$[{\boldmath $\alpha$}({\bf X})]}, for some \boldmath
$\alpha$, which is a set of functions of all the physical variables
${\bf X}$ on which mass loss rates do actually depend. In this way,
\boldmath $\alpha$ \unboldmath gives a mapping between the true
dependence of mass loss (including stochastic effects) and the
necessarily inadequate formulation we use here. For example, the
distribution $P(\etar)$ might arise from a stochastic distribution in
rotation rates, $\cal{P}(\omega)$. Even though we consider only simple
distributions for $P(\etar)$ we recognize that quite complex forms for
$P(\etar)$ could arise from a simple gaussian $\cal{P}(\omega)$.
Indeed, smooth variations in quantities like rotation and magnetic
fields often lead to bifurcated outcomes.

\begin{figure}[htbp]
\vbox to5.0in{\rule{0pt}{3in}}
\includegraphics{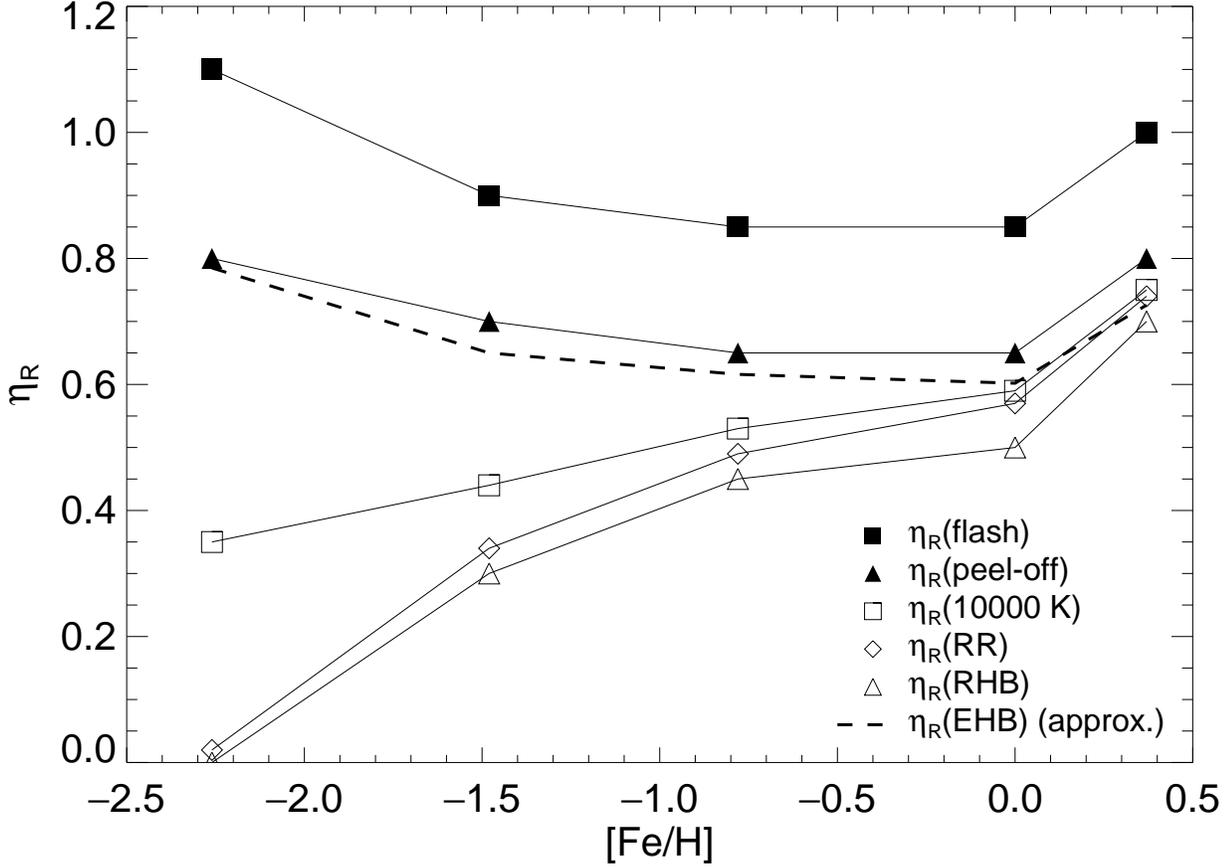}
\caption[feh vs eta]{\small Plot of $\etar$ vs Metallicity: The plot of
$\etar$ vs [Fe/H] shows $\etar$ that corresponds to (i) the blue end
of RHB --- $\etarhb$, (ii) RR Lyraes --- $\etarr$, (iii) effective
temperature of 10000~K --- $\etabhb$, (iv) $\etapeel$ --- the
transition between classic tip flashers and hot He flashers, (v)
$\etafl$ --- the largest value of $\etar$ for which a flash occurs.
The dashed line is the approximate location of \etaehb, the
lowest value of $\etar$ for which EHB stars form. The EHB stars very
roughly occupy the region between \etaehb\ and $\etafl$. Note the range
of $\etar$ corresponding to the mid-HB, $\etarhb - \etabhb$, decreases
as metallicity increases.
\label{fig:feheta}}
\end{figure}

We now will use our results to consider how the frequency of EHB
stars, or more specifically blue-hook stars, might vary with the
metallicity of a stellar population. We have plotted $\etar$ vs [Fe/H]
in Figure~\ref{fig:feheta} for $\etar$ values that correspond to
various HB fiducial points.
Using $\etar$ as the quantity on which the
distribution depends, we can crudely estimate how the ZAHB population
changes with metallicity. We define $\etarhb$ to be the largest value
of $\etar$ that still produces red HB stars at the ZAHB.  $\etarr$ is
the value of $\etar$ that marks the center of the RR Lyrae strip
(approximately $\logteff = 3.85$); $\etabhb$ (the value of $\etar$
that produces stars at 10,000 K) is roughly the hot end of the
``horizontal'' part of the HB.  The transition between the traditional
RGB tip-flashers and the hot-flashers is $\etapeel$. The largest value
of $\etar$ that allows a helium flash is $\etafl$.  The blue hook
stars lie between $\etapeel$ and $\etafl$, and the other EHB stars
very approximately occupy the region from $\etapeel - 0.05$ to
$\etapeel$. The approximate red end of the EHB population, \etaehb\ has
been indicated with the dashed line in Figure ~\ref{fig:feheta}. This
is a very crude estimate of where the cool end of the EHB population
should lie and was obtained by using the post-RGB tracks from DRO93.

There are a number of conclusions one can reach from this figure.
Throughout bear in mind that there are some EHB stars in globular
clusters, and in some, if not all, cases they originate from extreme
mass loss evolution in single stars. For example, there are many EHB
stars in \ocen, a low density, unrelaxed cluster with few classical
indications of stellar interactions such as blue stragglers (Bailyn
\etal\ 1992). The EHB stars are not concentrated toward the
center; thus even in an unfavorable environment for binary interactions,
many such stars are formed (see DOR95 for a discussion of single vs
binary origin of EHB stars).

In summary, note the following:

\begin{itemize}

\itm At low metallicity the range of values of $\etar$ which produces
EHB stars is comparable in magnitude to that producing the normal HB.
{\it Producing EHB stars requires no more fine tuning than producing
ordinary HB stars in globular clusters.}

\itm Despite the fact that high metallicity stars
must lose more mass to reach the EHB (for comparable age systems) they
can achieve this larger mass loss with \etar\ values that are
comparable to those needed for the low metallicity stars. Viewed in this way,
it is no more
difficult to produce EHB stars at high metallicity than at low, although
the corresponding physical interpretation of this result is unclear.

\itm The range of the mid-HB lies approximately from $\etarhb$ to
$\etabhb$. Figure~\ref{fig:feheta} shows that as metallicity
increases, this range decreases. It becomes more difficult to make
ordinary blue HB (IBHB) stars at higher metallicity---a point made by
DOR95 but more explicitly demonstrated here.  The ZAHB becomes
increasingly bimodal for metal rich populations. This is consistent
with observations of NGC 6791 which show a lack of IBHB stars and
three sdB (EHB) stars (Liebert \etal\ 1994).

\itm The range of values of $\etar$ producing EHB stars is almost constant
for all metallicities. To make EHB stars at high metallicity, \etar\
does not have to increase and its distribution does not have to be
more fine tuned.

\end{itemize}

\subsection{Distribution of Stars on the ZAHB} \label{subsec:ZAHBdist}

Considering the ZAHB population distribution as a function of
\etar\ instead of mass leads to a different perspective on
the production of EHB stars. The ZAHB mass depends almost linearly on
$\etar$ till one get close to the blue hook. For the blue hook stars
$M_{\rm ZAHB}$ does depend linearly on $\etar$ but with a slope that
is different from the rest of the ZAHB. The range in $\etar$ for the
blue hook is comparable to that of the rest of the ZAHB, but the range
in mass of the blue hook stars is much smaller than the mass range of
the ZAHB stars. Figure~\ref{fig:feheta} shows that the EHB is made up
of blue hook stars and slightly cooler extremely blue HB stars.  The
blue hook stars will contribute substantially to the EHB population if
the {\it support} of $P(\etar)$ (the range of \etar\ where
$P(\etar) \ne 0)$ is sufficiently great.
Therefore modeling the distribution of the ZAHB as a function of
$\etar$ ensures a reasonably well populated EHB.  This circumstance
arises jointly from the rapid increase in \mcore\ and the escalation
of the mass loss rate close to the RGB tip.  In some sense, ZAHB
models constructed with a $P(\etar)$ are more directly related to the
underlying physics of mass loss than those using $P(M_{\rm ZAHB}),$
since the star ``sees'' mass loss rates, rather than ``knows'' its
final ZAHB location.

\begin{figure}[htbp]
\vbox to5.0in{\rule{0pt}{3in}}
\includegraphics{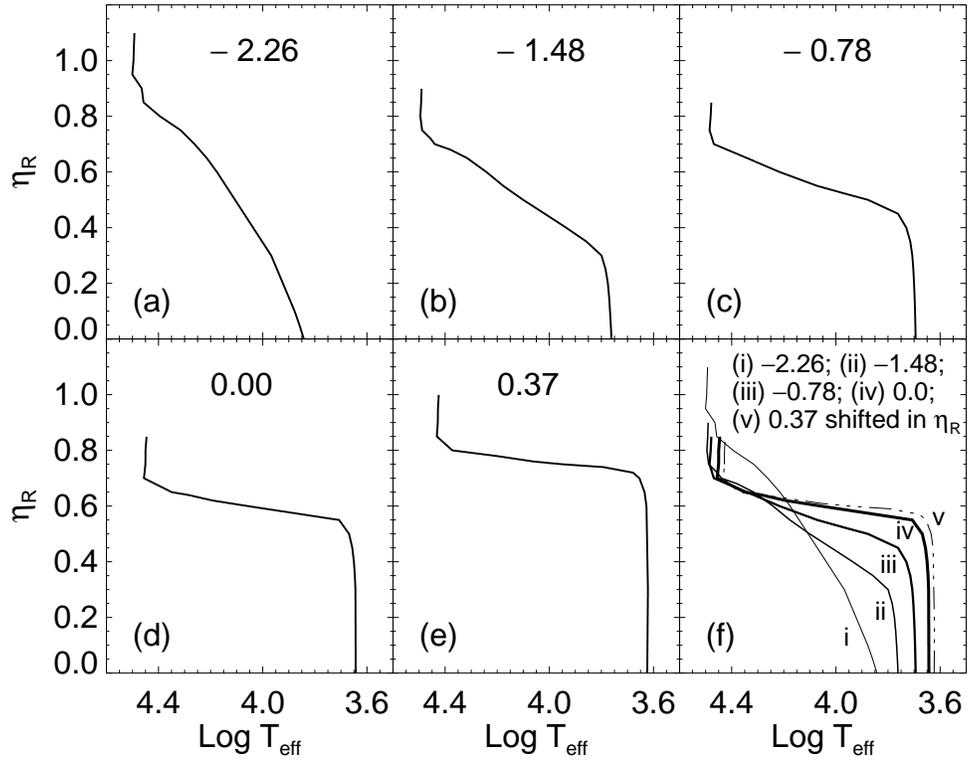}
\caption[eta vs temp]{\small Plot of $\etar$ vs \logteff\ for
the ZAHB stars plotted in Figure~\ref{fig:ZAHB}. The metallicity is
indicated at the top of each panel for panels (a) to (e). Panel (f)
shows all the plots for comparison with the $\feh = 0.37$ plot shifted
down by 0.15 in $\etar$ for clarity. \label{fig:etatem}}
\end{figure}

	Figure~\ref{fig:etatem} shows the relationship between $\etar$
and \logteff\ of the ZAHB stars for the five metallicities used in
this paper.
The slope, ${d\, \etar}/{d\, \logteff}$,
%$\frac{\textstyle d \etar}{\textstyle d \logteff}$,
is larger at the blue and red ends of the ZAHBs.
The slope of the mid-HB region is seen to decrease as metallicity
increases. This is more clearly shown in Figure~\ref{fig:etatem}(f)
where all the curves are superposed on each other. If the slope of
these curves is an indication of the number of stars present at a
given effective temperature on the ZAHB then it can be used to
describe the ZAHB distribution to first order. The number of stars per
\logteff\ interval, N(\logteff, \logteff +{\rm d}\,\logteff), will be given by

$$ N(\logteff,\,\logteff +{\rm d}\,\logteff) = P(\etar)\;
\frac{\textstyle d\, \etar}{\textstyle d\,\logteff}.$$

\begin{figure}[htbp]
\vbox to4.5in{\rule{0pt}{3in}}
\includegraphics{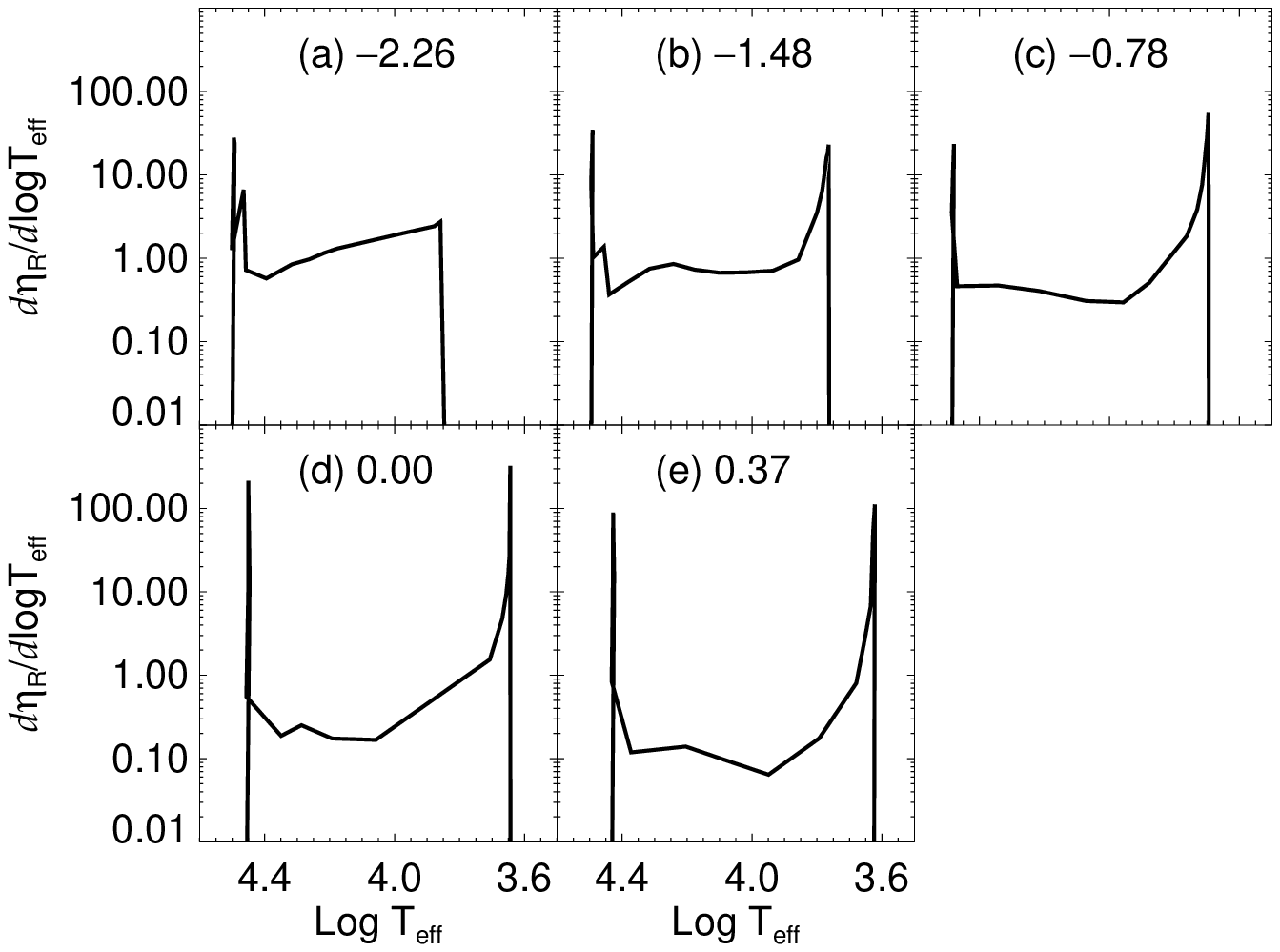}
\caption[detadt vs temp]{\small Plot of the slope, ${d\, \etar}/{d\,
\logteff}$,
of the curves in Figure~\ref{fig:etatem} vs \logteff . The slope is
directly related to the distribution of stars on the ZAHB for equally
probable $\etar$.  This is seen by comparing this figure and
Figure~\ref{fig:temhist}.  Note that these curves are not smooth
because of the discreteness of the data. The derivatives at the ends
are very large, hence the contrast between the ends and the mid-HB
portion is increased relative to the binned distribution shown in
Figure~\ref{fig:temhist}. The metallicity is indicated at the top of
each panel.
\label{fig:detadt}}
\end{figure}

\begin{figure}[htbp]
\vbox to5.0in{\rule{0pt}{3in}}
\includegraphics{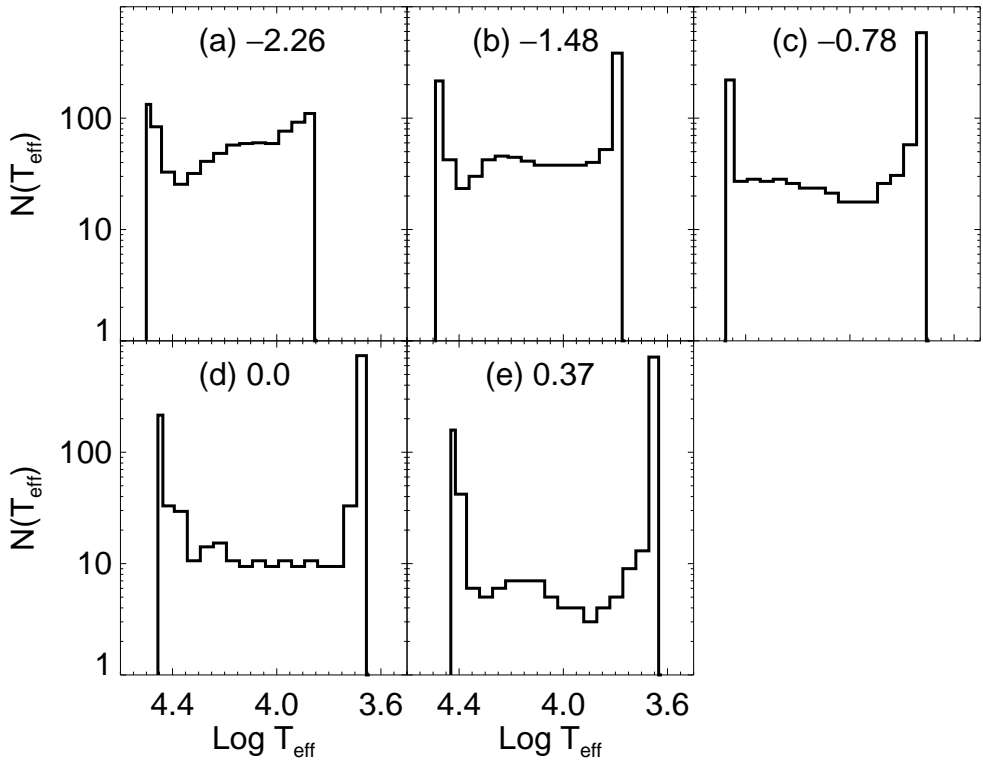}
\caption[temphist]{\small The effective temperature histogram of a thousand
ZAHB stars. The metallicity is indicated at the top of each panel.
The data are binned at intervals of 0.05 in \logteff . Note that these
are log plots, so the decrease in the number of mid-HB stars with
increasing metallicity appears somewhat exaggerated. \label{fig:temhist}}
\end{figure}

\noindent The normalized slope of these curves is shown in
Figure~\ref{fig:detadt}.
Note that these curves are not smooth because
of the discreteness of the numerical data used to construct them.  The
derivatives at the ends are very large.  Hence the contrast between
the ends and the mid-HB portion is increased relative to anything
which could be observed. Both observational error and the binning
required in a differential distribution function will smooth the ends.
Figure~\ref{fig:temhist} shows the effective temperature distribution
of a thousand ZAHB stars resulting from a flat distribution in $\etar$
i.e.~$P(\etar) = $ constant.  The flat distribution of a thousand
stars illustrated is created by linear interpolation of the ZAHB
models computed from the flashing RGB stars.  One can see that the
curves in Figure~\ref{fig:temhist} compare well with those in
Figure~\ref{fig:detadt}.

Using $P(\etar),$ both RHB and EHB stars can be produced if $P(\etar)
\ne 0$ in the appropriate range. When cast as a ``final'' mass
distribution $P(M_{\rm ZAHB})$ it appears that fine tuning is
required, but when $P(\etar)$ is used, it becomes clear that a fairly
wide range of mass loss behavior produces EHB stars.  However, for the
metal rich populations, fine tuning is necessary to populate the
mid-HB region.  The bimodal nature of the metal rich ZAHBs arises both
from the increased opacity of the metal rich stars and the increased
importance of the H (via CNO cycle) burning shell.  The envelopes of
EHB stars are so small that shell burning is negligible and the
opacity changes the temperature without affecting the structure
(Dorman 1992b). The temperature of the RHB is basically determined by
the position of the Hayashi line (fully convective stars). As
metallicity increases, both the opacity and H shell burning ``push''
stars from the mid-HB toward the Hayashi line, thus depopulating the
mid-HB.

These flat distributions are instructive because they give a measure
of the ``underlying'' effect of mass loss on the stellar mass
distribution as the rate is varied.  More realistic distributions as
found in some clusters  resemble gaussians, or at least are unimodal
with small
dispersion. It is easy to see that such dispersion functions will
produce sharply peaked mass distributions if the peak lies at a mass
corresponding to either high or low temperature.
Figure~\ref{fig:gauss} shows the temperature distributions that result
from gaussians of the same width but with different peak locations.
\begin{figure}[htbp]
\vbox to8.0in{\rule{0pt}{3in}}
\includegraphics{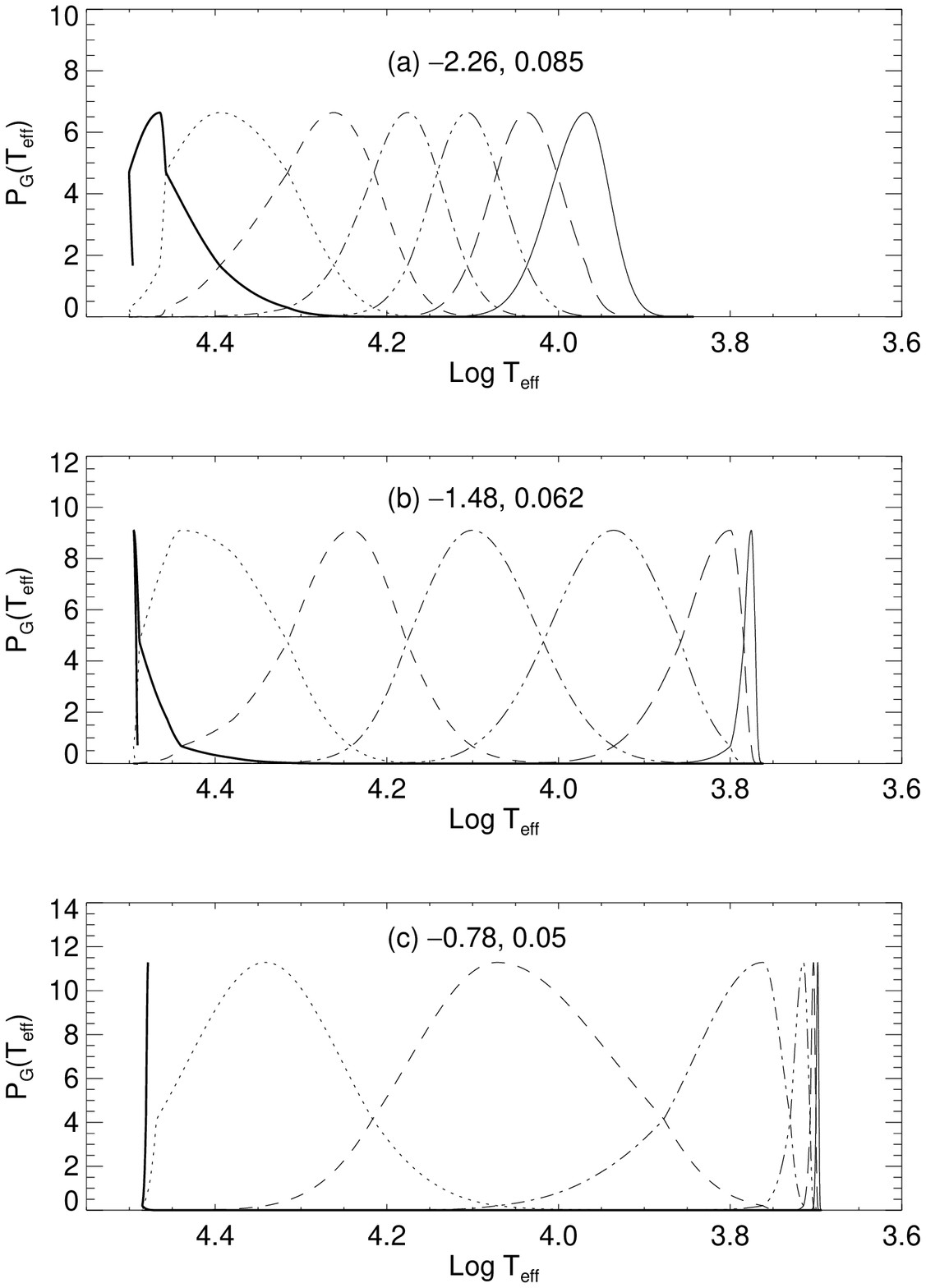}
\end{figure}
\begin{figure}[htbp]
\vbox to5.5in{\rule{0pt}{3in}}
\includegraphics{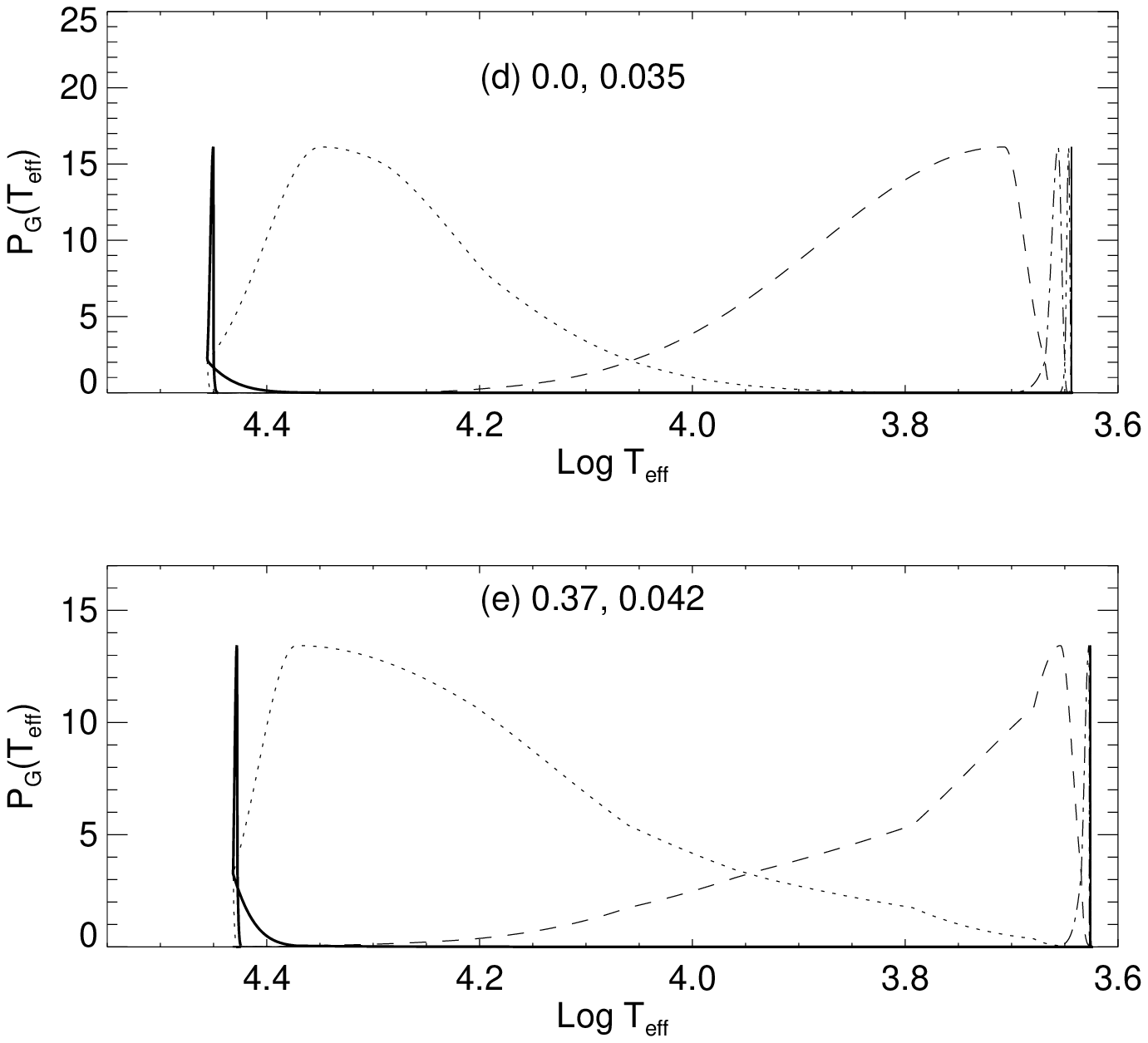}
\caption[Gaussian]{\small The distribution of the gaussian probability density,
$P_G(\teff)$, of the effective temperatures for the ZAHB stars. The
metallicity and the gaussian width in $\etar$, $\sigma(\etar)$, are
listed at the top of each panel.  The gaussian peaks are located at
$\etapeel +0.1, +0.0, -0.1, -0.2, -0.3, -0.4 -0.5$, going from left to
right. The width $\sigma(\etar)$ is set such that it gives a width of
0.03\msun\ in mass at $\etapeel -0.3$. Note that the gaussian
distributions change smoothly at low metallicities for different peak
positions. (The vertical lines at the ends of some of the plots are
very narrow distributions not fiducial marks.) At high metallicities
the distributions change sharply for peaks located in or near the
mid-HB region, but the distributions are identical when the peak lies
within the RHB. \label{fig:gauss}}
\end{figure}
The gaussian peaks were chosen relative to $\etapeel$ and are located
at $\etapeel +0.1, 0.0, -0.1, -0.2, -0.3, -0.4, -0.5$. The
$\sigma(\etar)$ values were chosen such that the width of the
corresponding gaussian in mass, $\sigma(M)$, is $0.03
\msun$ when the gaussian peak is located at $\etapeel - 0.3,$. We
choose a fiducial width of 0.03$\msun$ in mass because that is the
typical gaussian width of easily modeled globular cluster HBs. For the
three lower metallicities the distributions change smoothly as the
peak $\etar$ decreases. One can possibly expect to have both RHB and
EHB stars using a single Gaussian at these metallicities.

	At high $Z$, the ZAHB distribution is very dependent on the
location of the Gaussian peak especially if the peak is located within
or near the range of $\etar$ corresponding to the mid-HB region. This
is again a consequence of the small range of $\etar$ that corresponds
to mid-HB stars in metal rich populations. The ZAHB distribution is
insensitive to the Gaussian peak position if the peak is located
anywhere within the region corresponding to RHB stars for the high
metallicities. Hence at high metallicities one cannot tell the true
distribution of the ZAHB if only RHB stars are present. Also, for
both RHB and EHB stars to exist in metal rich populations, one has to
use a single gaussian which is very broad (much larger than estimated
for globular clusters) or use some kind of bimodal distribution.

%--------------------------SUMMARY---------------------------------

\section{SUMMARY} \label{sec:summary}

We have made an extensive study of hot-He-flash stars. These are stars
with such high mass loss rates that they peel-off the RGB before He
ignition, but rather than quietly becoming He white dwarfs, ignite He
while at high \teff. After He ignition the hot-flashers settle on a
blue-hook at the hot end of the HB and evolve as EHB stars (DOR95).

In the absence of an alternative, we used the Reimers mass loss
formula to compute red giant mass loss rates. We then consider the
distribution of initial HB masses as a probability distribution of the
Reimers' efficiency, \etar, rather than directly as a mass probability
distribution. The hot-flash behavior was surprisingly widespread.  For
globular cluster abundances the range of \etar\ producing EHB stars
was comparable to that producing normal HB stars. As metallicity is
increased, the range and magnitude of \etar\ leading to EHB stars
varies only slightly, whereas the range of \etar\ producing mid-HB stars
becomes very small. At solar metallicity and higher, the HB will be
composed of a group of very hot and very cool stars with little in
between so long as nature provides a broad enough distribution in
\etar.

It has been thought for some time that the upturn in the far-UV spectra
of elliptical galaxies and spiral bulges (Burstein \etal\ 1988) could
arise from the UV flux from EHB stars and their progeny (GR90;
\cite{RWO93}; DRO93; DOR95). The difficulty has been understanding the
origin of the EHB stars in metal rich populations. We have shown here
that it is just as easy to make EHB stars at high metallicity as at
low metallicity. Further, we do not require that the helium abundance
continue to grow as steeply with metallicity for $Z > Z_\odot$ as is
observed at low metallicity ($\Delta Y \sim 3 \Delta Z$) (Horch,
Demarque, \& Pinnsoneault 1992). In claiming that it is just as
easy to understand EHB stars in elliptical galaxies as in globular
clusters, we have, unfortunately, just passed the buck. We do not
really know why some stars in some clusters happen to lose 2--3 times
more mass than the typical star. But we do have the comfort of knowing
that such stars are observed and the hope that more detailed study of
these stars can lead to better understanding of stellar evolution and
stellar populations.

NLD thanks Jonathan Whitney for useful discussions.  This research was
supported by NASA Long Term Astrophysics Grant NAGW-2596, and by NASA RTOP
188-41-51-03.

%-----------------------------REFERENCES------------------------

\end{document}